\documentstyle[12pt,aasms4]{article}

\begin{document}

\title{A New Orbital Ephemeris and Reinterpretation of Spectroscopic Data
for the Supersoft X-ray Binary RX~J0513.9$-$6951\footnote{This paper
utilizes public domain data obtained by the MACHO Project, jointly funded
by the US Department of Energy through the University of California,
Lawrence Livermore National Laboratory under contract No.\ W-7405-Eng-48,
by the National Science Foundation through the Center for Particle
Astrophysics of the University of California under coopertative agreement
AST-8809616, and by the Mount Stromlo and Siding Spring Observatory, part
of the Australian National University. }} 

\author{A.P. Cowley\altaffilmark{2}, and P.C. Schmidtke\altaffilmark{2}}

\affil{Department of Physics \& Astronomy, Arizona State University,
Tempe, AZ, 85287-1504; anne.cowley@asu.edu, paul.schmidtke@asu.edu } 

\and

\author{D. Crampton and J.B. Hutchings}
\affil{Herzberg Institute of Astrophysics, NRC of Canada, Victoria, B.C.
V8X 4M6, Canada}

\altaffiltext{2} {Visiting Astronomers, Cerro Tololo Inter-American
Observatory, National Optical Astronomy Observatories, which is operated
by the Association of Universities for Research in Astronomy, Inc., under
contract with the National Science Foundation.} 

\begin{abstract}

We have analyzed nearly eight years of MACHO optical photometry of the
supersoft X-ray binary RX~J0513.9$-$6951 and derived a revised orbital
period and ephemeris.  Previously published velocities are reinterpreted
using the new ephemeris.  We show that the spectroscopic characteristics
of the system depend strongly on whether the system is in a high or low
optical state.  We also discuss the properties of the source's high/low
optical states and its long-term light curve.  Evidence for a 83.3-day
periodicity in the photometry is presented. 

\end{abstract} 


\section{Introduction}

The supersoft X-ray source RX~J0513.9$-$6951 (hereafter called X0513$-$69)
was discovered with the $ROSAT$ satellite (Schaeidt, Hasinger, \& Trumper
1993) and identified with a $\sim$16.7 mag emission-line star in the Large
Magellanic Cloud (Pakull et al. 1993; Cowley et al. 1993).  This star
appears to be the same as the variable HV 5682 found by Leavitt (1908). 
Its systemic velocity places it in the LMC, hence implying an absolute
magnitude of M$_V\sim-2.0$, the most luminous of all the supersoft X-ray
binaries (e.g. Cowley et al. 1998).  Long-term monitoring of X0513$-$69
shows that it is normally in a high optical state, but that it fades by
about a magnitude every 100-200 days, remaining in a low state for about a
month each time (e.g. Alcock et al. 1996). 

The optical spectrum shows extremely strong He II 4686\AA\ emission with a
narrow peak and extremely broad wings which are also weakly seen in the
Balmer and Pickering lines.  Weaker emission lines of O VI, N V, C IV, and
C III are also present.  Highly red- and blue-shifted ($\pm\sim$4000 km
s$^{-1}$) components of the strongest emission lines appear to come from
bi-polar jets (Crampton et al. 1996, Southwell et al. 1996).  The bright
absolute magnitude and the presence of jet emission lines suggest that the
accretion disk is very luminous and the system is undergoing rapid mass
transfer. 

In a spectroscopic study of X0513$-$69, Crampton et al.\ (1996) derived an
orbital period of $\sim$0.76 days from small radial velocity variations
(K$\sim$11 km s$^{-1}$) of the emission lines.  A similar photometric
period was found by Motch \& Pakull (1996) who showed the orbital light
curve has a single minimum and a full range of $\sim$0.06 mag.  An
improved photometric period was derived from three years of MACHO data by
Alcock et al.\ (1996) who found P=0.76278 days with a full amplitude of
$\sim$0.04 mag during the high optical state.  The very small photometric
and velocity amplitudes suggest the system is seen at a low orbital
inclination.  If the optical emission lines are assumed to show the
orbital motion of the compact star, the velocities imply the secondary is
an evolved low-mass ($<0.3M_{\odot}$) star. 

However, inconsistencies in the published ephemerides given by different
authors prompted us to use the entire MACHO data set to derive an improved
orbital period so that the existing optical data and new far-ultraviolet
observations (Hutchings et al. 2002) could be intercompared.  This new
period analysis is based on nearly 8 years of MACHO data.  Properties of
the orbital light curve, long-term photometric variations, and a
rediscussion of published spectroscopy are given in this paper. 

\section{Optical Light Curve}

\subsection{Long-term Changes in the Optical Light Curve}

The interactive MACHO Project light curve browser\footnote{
http://www.macho.mcmaster.ca/Data/MachoData.html} was used to extract
instrumental magnitudes for X0513$-$69.  We note that the MACHO website
displays light curves labeled ``blue" and ``red",  but after application
of the transformation equations given by Alcock et al. (1999), the colors
are close to $V$ and $R$, respectively.  Hereafter we will refer them as
$V_{MAC}$ and $R_{MAC}$.  As noted by many authors, the stellar image of
X0513$-$69 is blended with a faint ($V=19.1$, $B-V\sim+0.6$) field star
lying just $1.3^{\prime\prime}$ to the southeast.  Data from the online
website show that this field star was resolved on nearly all of the red
images but on none from the blue.  (Apparently, the field star is not
included in the blue image reduction template because of its rather red
color.)  Therefore, we had to explicitly remove an assumed contribution
of this contaminant from the $V_{MAC}$ values.  The $R_{MAC}$ magnitudes
could be used without correction.  Observations taken in only one color
or having an error $>0.05$ mag in either color were rejected from further
analysis.

The long-term $V_{MAC}$ light curve of X0513$-$69, covering MACHO
observations taken from 1992 through 1999, is displayed in Figure 1.  It
clearly shows the source undergoes semi-regular drops of about a magnitude
in its brightness level.  There were 16 low states which occurred during
the series of observations.  However, the minima are not equally spaced
and their depths and durations are not constant.  The separation of the
minima during the MACHO observation period ranged from 104 days to 195
days, with the average value being $\sim$161 days.  The depths (excluding
the deep dip at the start of the low states) ranged from $\sim$0.3 to
$\sim$1.2 mag in $V_{MAC}$, with the average being $\sim0.64$ mag.  There
appears to be a gradual trend for the minima to become shallower from 1992
to 1999, although the data are sparcer after 1997 so the minima are not as
well defined.  Also, the duration of the minima increased slightly from
$\sim$35 days at the start of the MACHO observations to $\sim$50 days by
the end of the data set.  There appears to be a very weak trend for the
magnitude at the start of each bright state to increase with time,
changing by $\sim$0.3 mag over eight years.  All of these factors need to
be considered if one is trying to extract the small orbital variations. 

\subsection{Revised Orbital Ephemeris from High-State Photometry}

In order to be able to compare new far-ultraviolet data with the older
optical data, we re-examined the published orbital periods and ephemerides
for X0513$-$69 (e.g. Alcock et al.\ 1996, Crampton et al.\ 1996, Motch \&
Pakull 1996, Southwell et al.\ 1996).  Although all authors give orbital
periods near $\sim0.76$ days, none of the studies is completely consistent
with the others, and each has some limitations.  For example, the
spectroscopic studies by Crampton et al.\ and Southwell et al.\ were based
on observations that covered only a few days in each observing season, so
that it is not possible to carry their phasing forward for many cycles. 
Similarly, the photometric data of Motch \& Pakull only covered 6 nights,
so the period was not well established.  Alcock et al.\ derived a
photometric period based on the first three years of MACHO data (1992 to
1995).  However, the ephemeris they presented is not consistent with that
given by Motch \& Pakull, although their dates of observation overlapped. 
These problems lead us to analyze the entire MACHO data set. 

Using the corrected $V_{MAC}$ magnitudes, we first removed all of the
observations when X0513$-$69 was in its low optical state or in a
transition between states, resulting in 720 usable $V_{MAC}$ magnitudes.
We then ``detrended" the data for each high-state interval by subtracting
a linear fit to the $V_{MAC}$ magnitudes, since the source consistently
shows a gradual fading of 0.1--0.3 mag during bright episodes.  A similar
procedure was followed by Alcock et al.  (We also tested this method using
a quadratic fit to the high-state data, but this did not change any of the
final results.)  To search for low-amplitude orbital modulations, a
periodogram of the residuals, ${\Delta}V_{MAC}$, was calculated (see
Figure 2).  There is a significant peak at P=0.76295 days, near the value
given by Alcock et al.  We refined the period by performing a
least-squares fit of the residuals to a sine curve which resulted in the
following $V_{MAC}$ ephemeris: 

T(min) = JD 2,448,858.099$\pm0.001$ + 0.7629434$\pm0.0000020$E 
                  ~~(= MJD 48857.599)

\noindent
Since data from the MACHO website are tabulated in Modified Julian Days
(MJD) we give the time of minimum light in both JD and MJD to indicate
that we have used the appropriate conversion.  The semi-amplitude of the
best-fit sine curve to the ${\Delta}V_{MAC}$ data is 0.021 mag.  Subsets 
of the data were also examined to look for possible period changes, but 
no significant differences were found.

A similar procedure was followed using the $R_{MAC}$ data.  We found a
formal period that differed by only slightly from the $V_{MAC}$ value and
a fitted sine curve that had a similar amplitude and time of minimum
light.  Therefore, we have adopted the $V_{MAC}$ ephemeris for the
subsequent analysis of all photometric and spectroscopic data.  In Figure
3 (upper panel) we plot the bright-state ${\Delta}V_{MAC}$ and
${\Delta}R_{MAC}$ light curves, with individual data points averaged into
30 bins for each color.  Error bars are only shown for one color, since
they would otherwise overlap each other.  The agreement between
${\Delta}V_{MAC}$ and ${\Delta}R_{MAC}$ curves in both phase and amplitude
is extremely good, and it demonstrates that there is no significant color
change through the orbit when the source is in its high optical state. 

\subsection{Low-State Optical Light Curve} 

As shown in Figure 1, X0513$-$69 drops from its normally high optical
state ($V_{MAC}\sim$16.7) by $\sim0.7$ mag quasi-periodically.  Leavitt
(1908) noted this star (HV 5682) varied by $\sim$1 mag between 1893 and
1906, so we know this system has been undergoing these changes for over
100 years.  It has been shown that during the optical dips the source
undergoes soft X-ray outbursts (e.g. Reinsch et al. 1996).  It is thought
that mass flow rate at the surface of the while dwarf is at a minimum
during the low optical state, while the bright states occur in times of
rapid flow rate (Reinsch et al. 2000). 

To analyze the faint-state data, we eliminated all observations made
during the high and transition states.  In addition, the short, deep dips
at the beginning of the low states were also removed.  During each minimum
the source tends to brighten slightly, so we have removed this trend for
each individual segment.  Only those intervals with a minimum of four data
points were analyzed.  Period analysis of the 145 faint-state residuals
yields a value very close to that found for the bright state.  In Figure 3
(lower panel) we plot the ${\Delta}V_{MAC}$ and ${\Delta}R_{MAC}$ light
curves for the low-state data using the bright-state $V_{MAC}$ ephemeris. 
Both faint-state light curves have a minimum at phase zero, a small
magnitude range, and no orbital color variation. 

\subsection{Comparison of High and Low Optical State Light Curves and Colors}

Figure 3 shows the light curves in both colors and from both the high and
low states are very similar in amplitude and phasing.  We emphasize that
these curves represent the orbital modulation of X0513$-$69 after removing
the mean light level.  The MACHO data were also examined to see if there
is an overall color change between the high and low states.  We find
($V-R$)$_{MAC}$=0.146 in the bright state and ($V-R$)$_{MAC}$=0.161 in the
low state, which indicates there is no significant change in color between
states.  Although Reinsch et al.\ (1996) found the system became slightly
redder [$\Delta$($B-V$)$\sim0.1$] during a single low state they observed
in 1994 December, their discussion suggests that the contribution of the
nearby field star may not have been removed.  The fairly red color of the
field star ($B-V=0.6$) would change the observed color of X0513$-$69 by
about what they observe if the extraneous contribution were not
subtracted. 

It is clear that irregular fluctuations are superimposed on the orbital
variations.  Thus, when all the individual MACHO residual magnitudes
(${\Delta}V_{MAC}$ or ${\Delta}R_{MAC}$) are plotted, the result is a band
of points covering a range of $\sim$0.2 mag at all phases.  The binned
light curve shown in Figure 3 becomes apparent only through long-term
monitoring.  If only a few cycles are observed, the resulting light curve
can differ considerably from the mean derived from many years of data. 
This is presumably why Motch \& Pakull's analysis of only 6 nights of data
gave a somewhat shorter period (0.745 days) than the extensive MACHO data
set.  Similar irregular behavior from cycle to cycle is typical in
cataclysmic variables and related objects. 

\subsection{Long-term Periodicities}

We also examined additional peaks in the periodogram that were not were
not simple aliases of the orbital period (see the upper panel of Figure
4).  A peak at P=83.2 days stands out in the high-state ${\Delta}V_{MAC}$
data.  Fitting the data to a sine curve, the full amplitude is $\sim$0.03
mag with a minimum at MJD 48889.53$\pm$0.16.  In the lower panel of Figure
4, we show the ${\Delta}V_{MAC}$ light curve of X0513$-$69 plotted on
P=83.2 days. 

We tested the significance of this periodicity in three ways.  In order to
be sure that the long-term period wasn't in some way related to the
orbital modulation, we removed the 0.76-day variation and recomputed the
periodogram.  The long-period signal remained after the orbital variations
were taken out.  We then tried an alternative detrending algorithm using a
quadratic rather than linear model.  The resulting periodogram still
showed a strong low-frequency peak, albeit at a slightly shorter period
(P=76.4 days).  Finally, we evaluated the significance of the long-term
periodicity by randomly assigning each of the ${\Delta}V_{MAC}$ values to
one of the observation times and recalculating the periodogram.  Since the
observed power of the 83.2-day period was exceeded in {\it none} of 250
random trials, the signal detected at this period -- which is coherent
over 8 years of data or about 35 cycles -- is highly significant. 

Using the low-state data, there is little evidence of this long period in
either the calculated periodogram or in the folded light curve.  Our
speculation is that the long period might be related to the disk's
precession. In the low state, the disk would be sufficiently faint that
its variation might not be detectable.

\subsection{CTIO Photometry of X0513$-$69}

We have obtained photometry of X0513$-$69 at Cerro Tololo Inter-American
Observatory during several observing runs, of which most were coordinated
with spectroscopic observations.  Our previously published photometry
(Cowley et al.\ 1996) includes data taken in 1992 December and 1993 March
when the source was bright and in 1993 December when X0513$-$69 was in a
transition state, $\sim$0.3 mag below its bright level.  Additional
bright-state data from 3 nights in 1994 November was published by Crampton
et al.\ (1996).  A single $V$ observation from 1995 November showed
X0513$-$69 in a bright optical state (Cowley et al.\ 1998). 

The new CCD photometry presented here was obtained with the CTIO 0.9 m
telescope on 1999 March 21--28 (UT), using the same instrumentation as in
1994.  Altogether, 14 pairs of $V$ and $B$ images were taken; they were
reduced and calibrated with procedures consistent with the processing
applied to the older data sets.  In particular, the faint field star was
resolved on each image taken in 1999.  In Table 1 we present the
previously unpublished $B$ and $V$ photometry, which indicate the source
was in a bright optical state.  The mean color of X0513$-$69 was
$B-V=-0.04$ during the observing run. 

The light curve from the 1999 March CTIO data is entirely consistent with
the MACHO data covering the same epoch.  Although there are not enough data
points in the small sample to determine an accurate period, the detrended
and folded CTIO light curve shows orbital modulations that lie within the
envelope of the MACHO bright-state data presented in $\S$2.2. 

\section{Reanalysis of Published Optical Spectroscopy} 

With a greatly improved ephemeris it should be possible to phase together
the radial velocities from previously published data.  We have included
data taken by Crampton et al.\ in 1992 December (Dec92), 1993 December
(Dec93), and 1994 November (Nov94) as well as 1994 November-December
(Nov/Dec94) spectra reported by Southwell et al.  A plot of these
velocities using the new ephemeris shows that these data are not well fit
by a single velocity curve.  The Nov/Dec94 and Dec93 velocities agreed
with each other in amplitude and phasing.  Similarly, the Dec92 and Nov94
data gave velocity curves consistent with each other, but differing from
the first one.  In trying to understand this, we examined the MACHO
long-term light curve to see what optical state the source was in during
each of these observations.  The Dec92 and Nov94 spectra were taken when
X0513$-$69 was in its bright state.  The Nov/Dec94 spectra were taken as
the source was transitioning into a low state.  Although there is a gap in
the MACHO data during the Dec93 spectroscopic observations, simultaneous
photometry obtained by Schmidtke (Cowley et al. 1996) show X0513$-$69 at
$V\sim17.0$, which is fainter than the bright state and consistent with 
the source being in transition to or from a low state. 

The velocity curves from the bright and intermediate/transition states are
shown in Figure 5.  When the source was in its high state, maximum
positive velocity was observed at phase $\sim$0.1.  When the source was in
an intermediate state, maximum velocity occurred about 0.2P earlier and
with a slightly higher amplitude.  This suggests that in the bright state,
when the mass transfer is high, the emission lines are strongly dominated
by the flow from the donor star so that maximum positive velocity is
observed near the conjuction at phase zero.  As the source transitions
into a lower state, the mass transfer is less and the accretion disk
around the compact star now contributes more to the emission lines, so
that some of the orbital motion of that star is evident (hence the maximum
velocity occurs nearer quadrature when the compact star's motion is away
from the observer).  We note that the systemic velocities in the two plots
appear to differ by $\sim$20 km s$^{-1}$.  We have made every effort to
ensure the Crampton et al.\ velocities are all on the same velocity system
and measured in a uniform way.  However, we have no way of knowing how the
calibrations for the Southwell et al.\ data were carried out.  Since the
Southwell et al.\ velocities comprise the majority of data points in the
intermediate state, a difference in their velocity scale or method of
measurement may cause the change in systemic velocity between the two
plots.  Further observations will be needed to see if the differences in
velocity curves between the two states are real or a calibration
uncertainty. 

We would expect that velocities obtained during a low state, when the mass
transfer is very low, would come mainly from the region near the compact
star.  That would be the ideal time to obtain a velocity curve with the
goal of better defining the stellar masses in this system.  Knowledge of
how the emission-line behavior changes depending on the brightness state
of the system should help us to interpret future spectroscopic
observations, both optical and ultraviolet. 

In summary, we have derived a new ephemeris for X0513$-$69 which allows us
to intercompare published photometric and spectroscopic data and to
understand how the system's characteristics change with brightness level. 

\acknowledgments

APC acknowledges her support from the National Science Foundation.  We
thank the referee for helpful and constructive comments.

\clearpage

\begin{deluxetable}{cccccc}
\tablenum{1}
\footnotesize
\tablecaption{1999 March Photometry of RX~J0513.9$-$6951}
\tablehead{
\colhead{HJD} &
\colhead{Phase\tablenotemark{a}} &
\colhead{$V$} &
\colhead{HJD} &
\colhead{Phase\tablenotemark{a}} &
\colhead{$B$} \\
\colhead{2450000+} & & &  
\colhead{2450000+} & 
}
\startdata

1258.5209 & 0.264 & 16.643$\pm$0.014 & 1258.5260 & 0.271 & 16.598$\pm$0.016 \nl
1259.5700 & 0.640 & 16.592$\pm$0.008 & 1259.5747 & 0.646 & 16.547$\pm$0.007 \nl
1259.5897 & 0.665 & 16.608$\pm$0.007 & 1259.5940 & 0.671 & 16.588$\pm$0.025 \nl
1260.5697 & 0.950 & 16.655$\pm$0.014 & 1260.5741 & 0.955 & 16.713$\pm$0.025 \nl
1260.5884 & 0.974 & 16.679$\pm$0.024 & 1260.5927 & 0.980 & 16.634$\pm$0.032 \nl
1261.5682 & 0.258 & 16.658$\pm$0.009 & 1261.5727 & 0.264 & 16.595$\pm$0.009 \nl
1261.5869 & 0.283 & 16.644$\pm$0.011 & 1261.5912 & 0.288 & 16.571$\pm$0.017 \nl
1262.5049 & 0.486 & 16.634$\pm$0.020 & 1262.5091 & 0.492 & 16.611$\pm$0.009 \nl
1262.5537 & 0.549 & 16.673$\pm$0.007 & 1262.5583 & 0.556 & 16.611$\pm$0.012 \nl
1262.5733 & 0.575 & 16.673$\pm$0.010 & 1262.5775 & 0.581 & 16.722$\pm$0.026 \nl
1263.5422 & 0.846 & 16.662$\pm$0.011 & 1263.5466 & 0.852 & 16.629$\pm$0.015 \nl
1263.5620 & 0.872 & 16.674$\pm$0.026 & 1263.5668 & 0.878 & 16.665$\pm$0.022 \nl
1264.5319 & 0.143 & 16.666$\pm$0.013 & 1264.5366 & 0.149 & 16.602$\pm$0.012 \nl
1265.5430 & 0.468 & 16.714$\pm$0.013 & 1265.5476 & 0.474 & 16.692$\pm$0.029 \nl
\enddata

\tablenotetext{a}{T$_0$ = HJD2448858.099 + 0.7629434E days, 
where T$_0$ is time of minimum light}

\end{deluxetable}

\clearpage

\begin{figure}
\caption{Long-term MACHO $V_{MAC}$ light curve of RX~J0513.9$-$6951
showing its high and low optical states. } 
\end{figure}

\begin{figure}
\caption{Power spectrum for the high-state MACHO ${\Delta}V_{MAC}$ data,
based on 720 data points, as described in the text.  The peak frequency
corresponds to a period of P = 0.7629434 days. } 
\end{figure}

\begin{figure}
\caption{(upper) ${\Delta}V_{MAC}$ (filled circles) and ${\Delta}R_{MAC}$
(open circles) light curves for RX~J0513.9$-$6951 in its high optical
state, folded on the newly derived period of P = 0.7629434 days and T(min)
= JD 2,448,858.099.  The data have been averaged in 30 phase bins. 
~(lower) ${\Delta}V_{MAC}$ and ${\Delta}R_{MAC}$ light curves from the
low-state data, using the same symbols for each color as in the upper
panel.  Because there are considerably fewer observations taken in the low
state, the data have been averaged into 15 phase bins. } 
\end{figure}

\begin{figure}
\caption{(upper) Low-frequency power spectrum for the high-state
${\Delta}V_{MAC}$ data, showing significant power at P=83.2 days. 
~(lower) ${\Delta}V_{MAC}$ (filled circles) and ${\Delta}R_{MAC}$ (open
circles) photometry plotted on the possible long-term period of P=83.2
days.  See text for further discussion. } 
\end{figure}

\begin{figure}
\caption{Velocity curves for He~II 4686\AA\ taken from Crampton et al.\
and from Southwell et al.  The upper panel displays the velocities
obtained when the source was in a bright optical state (crosses for 1992
December, squares for 1994 November), while the lower panel shows
observations obtained when the source was in an intermediate or
``transition" state (circles for 1993 December, pluses for 1994
November/December).  Note how the phasing changes between the two states. 
} 
\end{figure}


\begin{references}

\reference{} Alcock, C., et al. 1996, MNRAS, 280, L49

\reference{} Alcock, C., et al. 1999, PASP, 111, 1539

\reference{} Cowley, A.P., Schmidtke, P.C., Crampton, D., \& Hutchings, 
J.B. 1996, in ``Compact Star in Binaries", ed. J. van Paradijs, E.P.J. van 
den Heuvel, \& E. Kuulkers, (Kluwer Academic Publishers), p. 439

\reference{} Cowley, A.P., Schmidtke, P.C., Crampton, D., \& Hutchings, J.B.
1998, ApJ, 504, 854

\reference{} Cowley, A.P., Schmidtke, P.C., Hutchings, J.B., Crampton, D.,
\& McGrath, T.K. 1993, ApJ, 418, L63 

\reference{} Crampton, D., Hutchings, J.B., Cowley, A.P., Schmidtke, P.C.,
McGrath, T.K., O'Donoghue, D., \& Harrop-Allin, M.K. 1996, ApJ, 456, 320 

\reference{} Hutchings, J.B., Winter, K., Cowley, A.P., Schmidtke, P.C., 
\& Crampton, D. 2002, AJ, submitted 

\reference{} Leavitt, H.S. 1908, Harvard Ann., 60, 87

\reference{} Motch, C. \& Pakull, M.W. 1996, Lecture Notes in Physics,
472, 127, in Supersoft X-ray Sources, ed. J. Greiner 

\reference{} Pakull, M.W., Motch, C., Bianchi, L., Thomas, H-C., Guibert,
J., Beaulieu, J.P., Grison, J.P., \& Schaeidt, S. 1993, A\&A, 278, L39 

\reference{} Reinsch, K. van Teeseling, A., Beuermann, K. \& Abbott, 
T.M.C. 1996, A\&A, 309, L11

\reference{} Reinsch, K., van Teeseling, A., King, A.R., \& Beuermann, 
K. 2000, A\&A, 354, L37

\reference{} Schaeidt, S., Hasinger, G., \& Trumper, J. 1993, A\&A, 270, L9

\reference{} Southwell, K.A., Livio, M., Charles, P.A., O'Donoghue, D., \& 
Sutherland, W.J. 1996, ApJ, 470, 1065

\end{references}
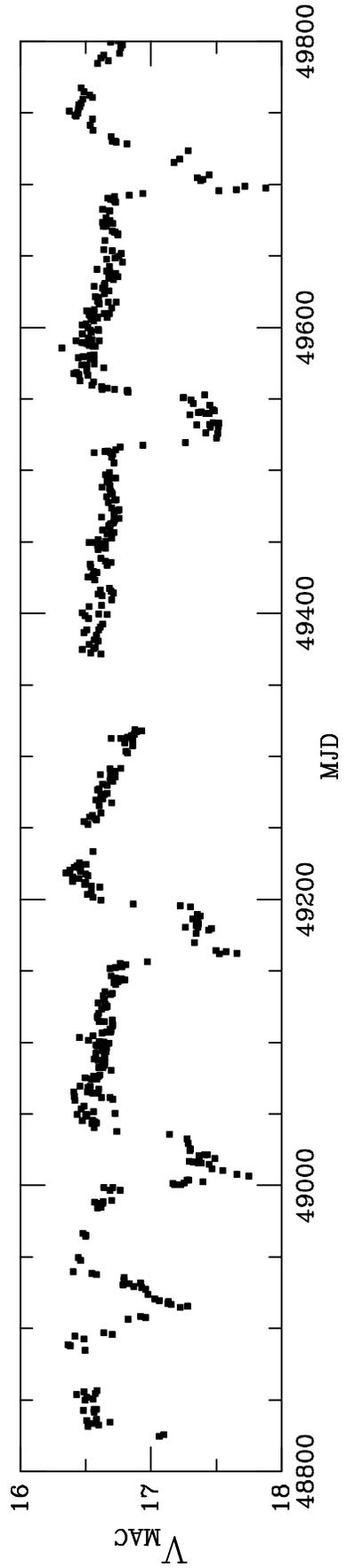
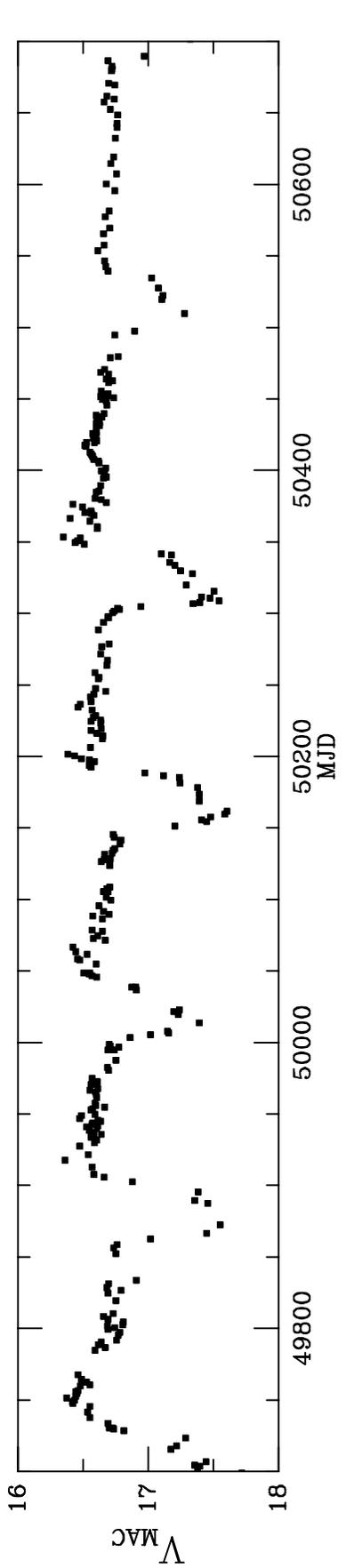
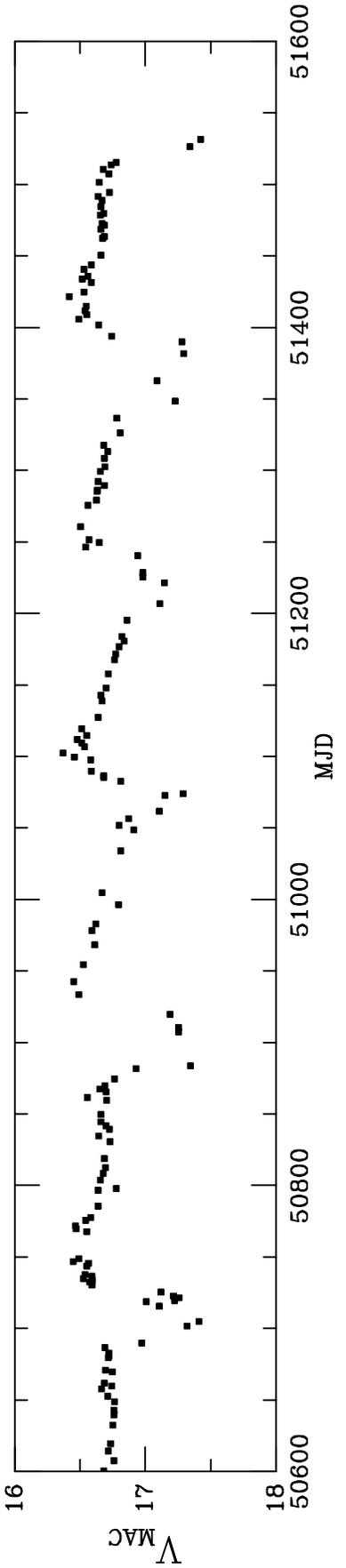
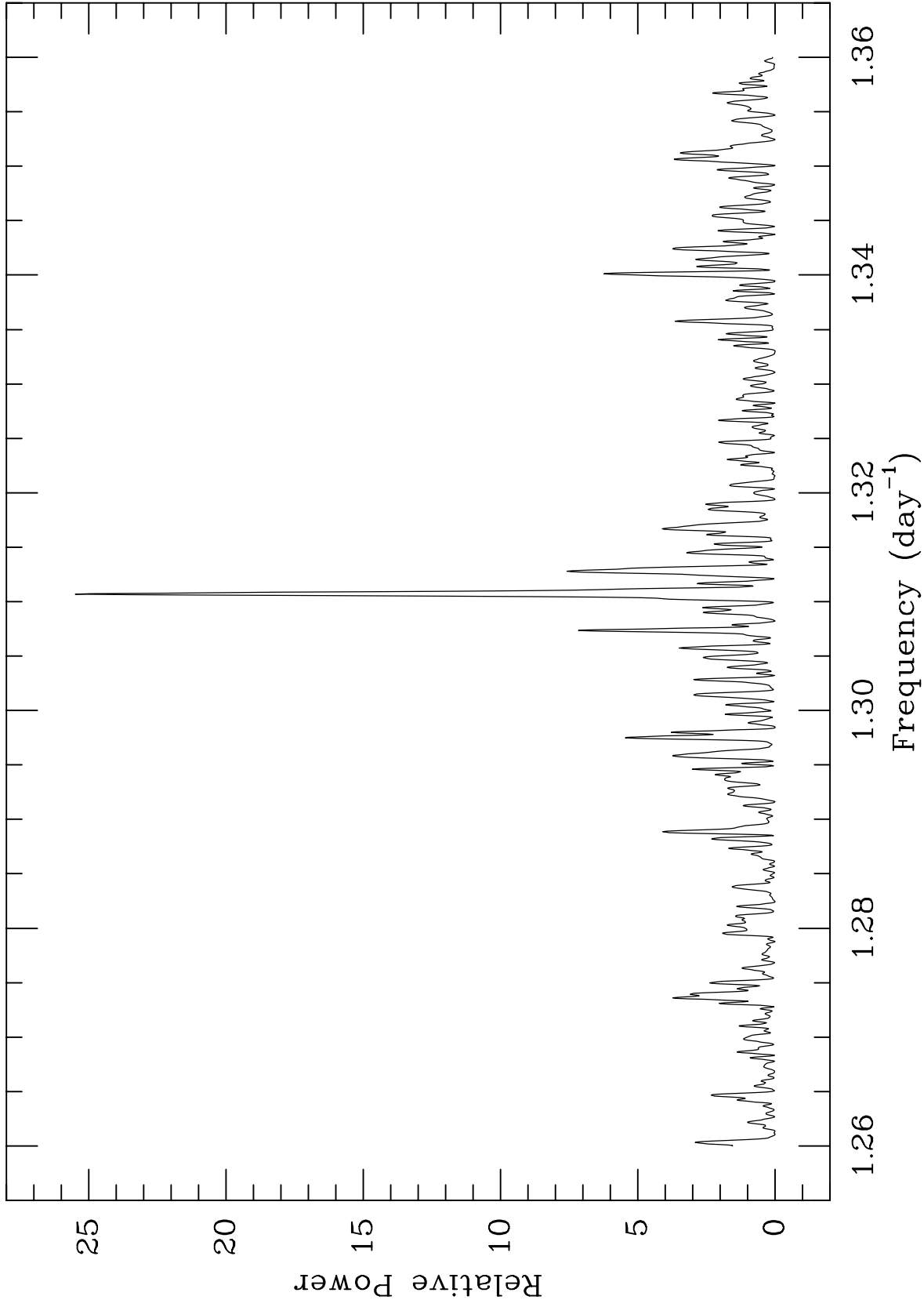
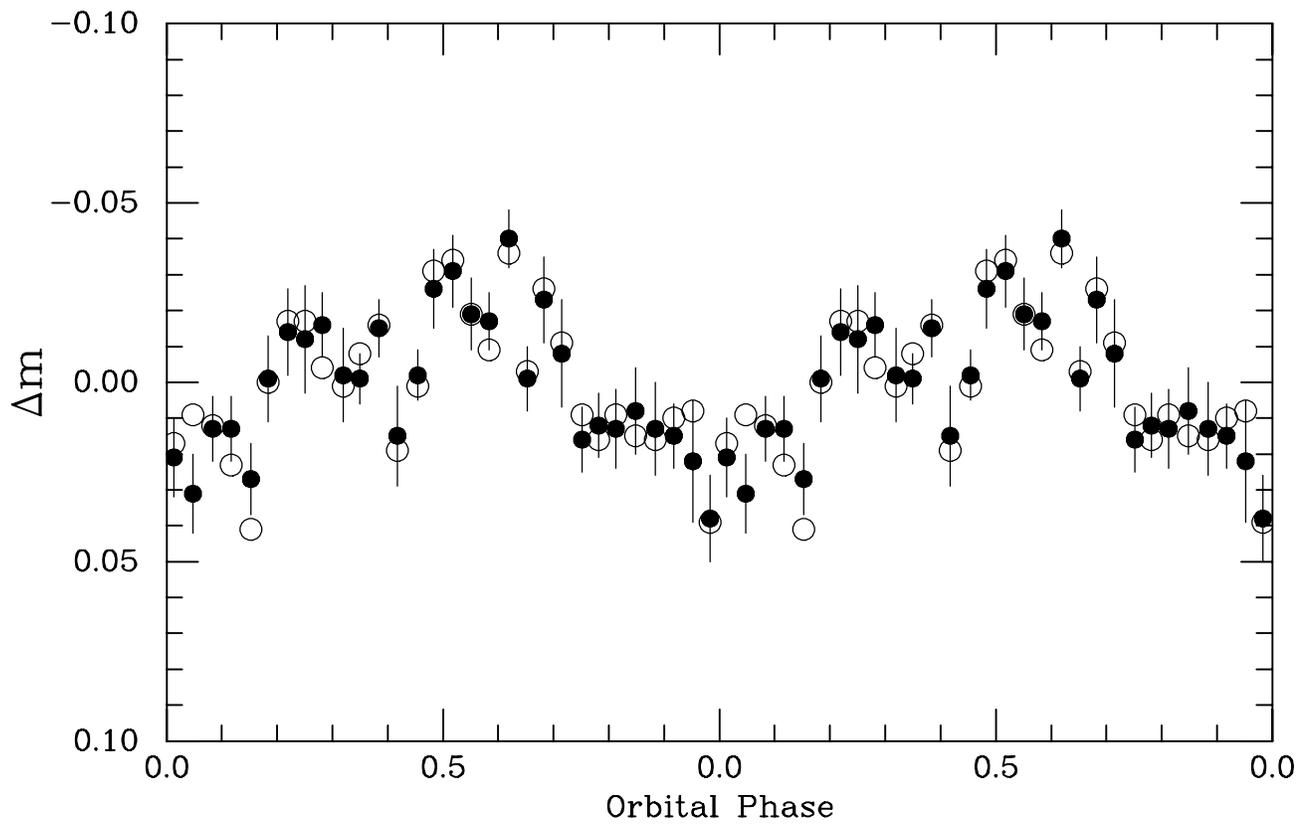
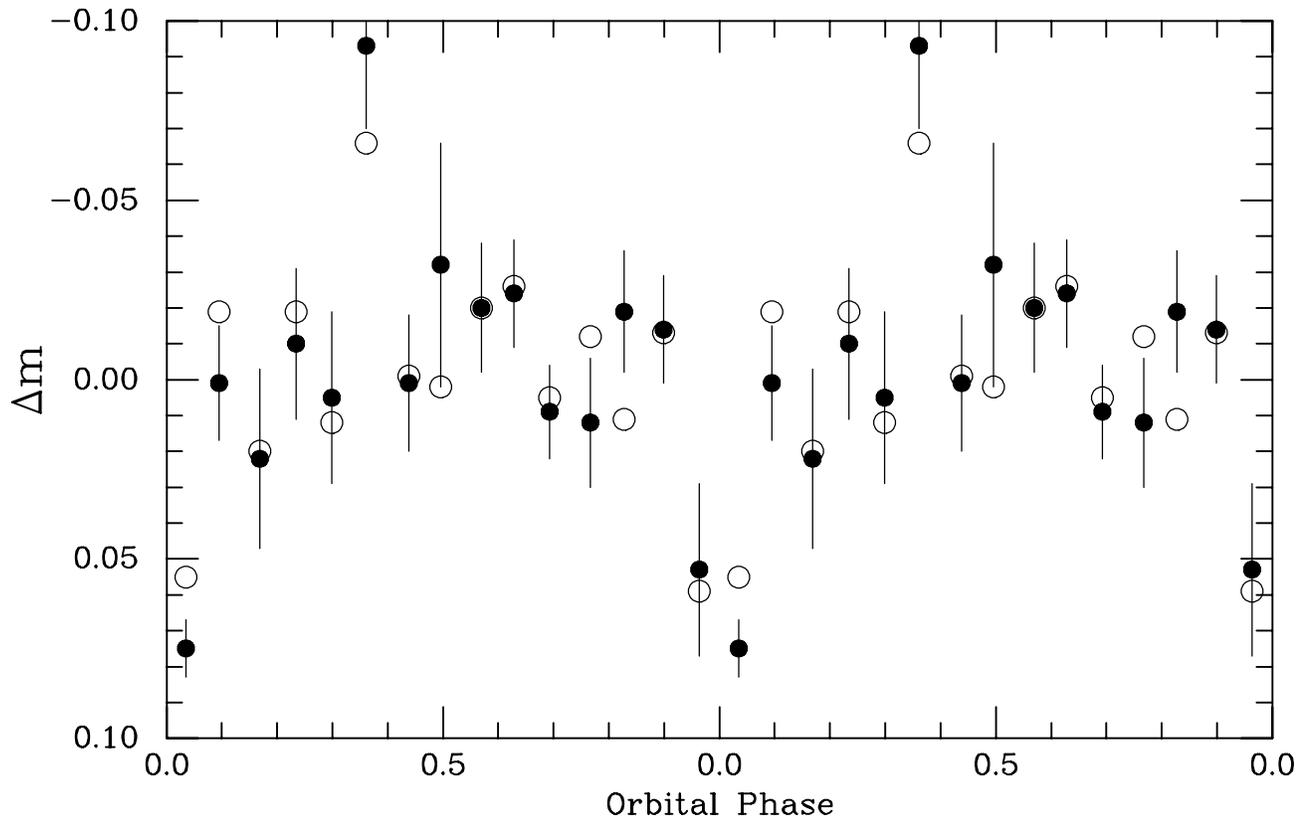
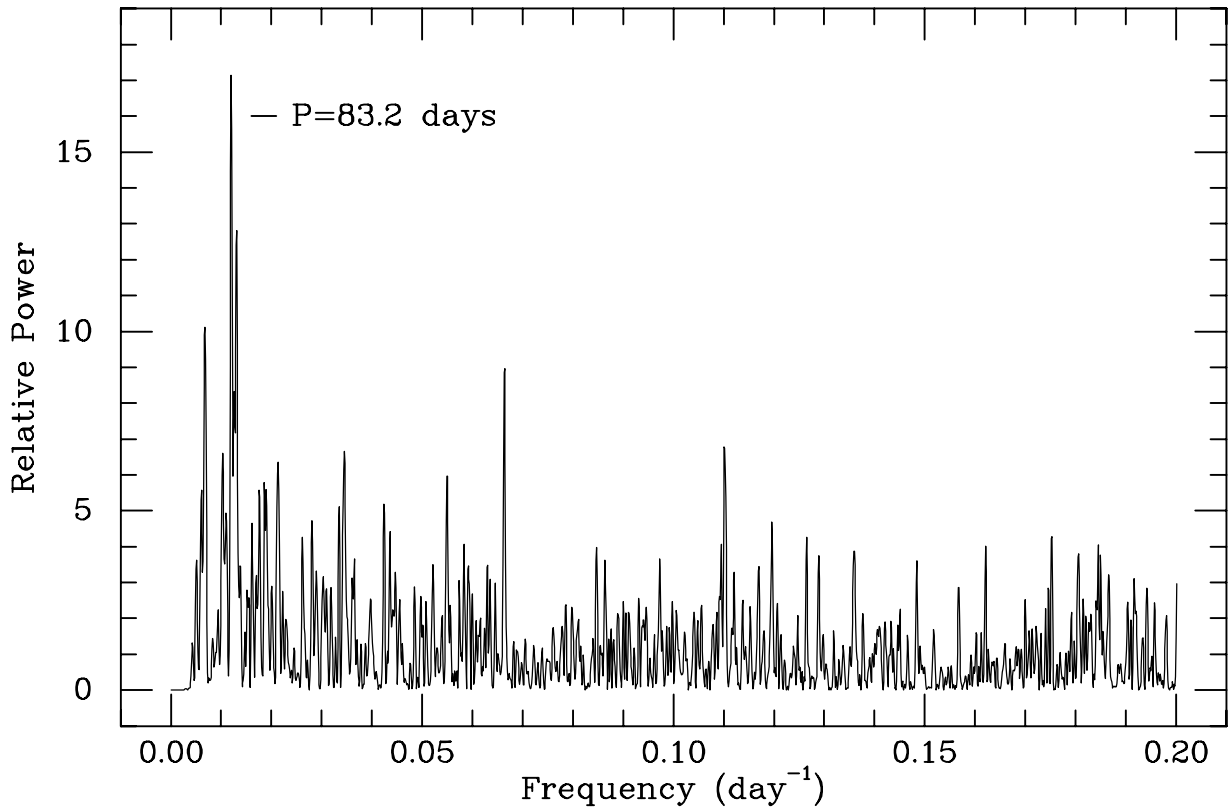
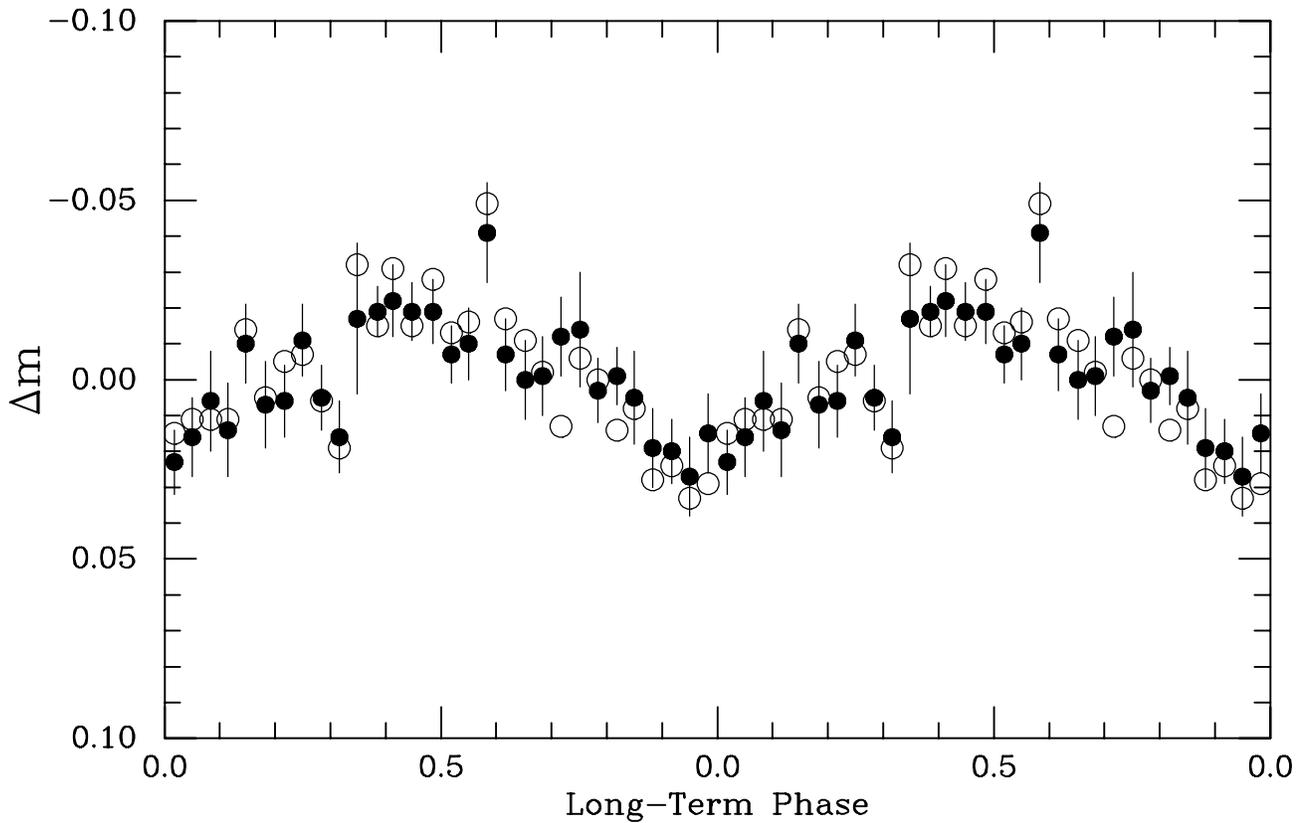
\end{document}